\documentclass[9pt, table, conference]{IEEEtran}
\IEEEoverridecommandlockouts
\usepackage{amsmath,graphicx}
\usepackage{cite}
\usepackage[utf8]{inputenc}
\usepackage{amsfonts}
\usepackage[spanish, english]{babel}
\usepackage{tcolorbox}
\usepackage{amsthm}  % For theorem and proof environments
\usepackage{mathrsfs}
\usepackage{xparse}
\usepackage{algorithm}
\usepackage{algpseudocode}
\tcbuselibrary{skins}
\usepackage{color}
\usepackage{makecell, cellspace}
\usepackage{setspace}
\usepackage{graphicx}
\usepackage{multirow}
\definecolor{Pastel}{RGB}{245,245,250}
\definecolor{Pastel2}{gray}{0.99}
\usepackage{verbatim}
\makeatletter
\newcommand*\bigcdot{\mathpalette\bigcdot@{.5}}
\newcommand*\bigcdot@[2]{\mathbin{\vcenter{\hbox{\scalebox{#2}{$\m@th#1\bullet$}}}}}
\makeatother
\usepackage{bibentry}
\usepackage{makecell, cellspace}
\usepackage{booktabs}
\usepackage[normalem]{ulem}
\usepackage{mathtools, physics}
\usepackage{hyperref}
\usepackage{cleveref} 
\crefname{section}{Sec.}{Secs.} % For singular "Sec. IV" and plural "Secs. I and II"
\crefname{subsection}{Sec.}{Secs.}
\crefname{subsubsection}{Sec.}{Secs.}
\crefname{figure}{Fig.}{Figs.}
\crefname{table}{Table}{Tables}
\crefname{equation}{Eq.}{Eqs.}
\crefname{algorithm}{Alg.}{Algs.} % If you use the algorithm environment

\newtheoremstyle{tight}
  {0.25em}   % Space above
  {0.25em}   % Space below
  {\itshape}  % Body font
  {}       % Indent
  {\bfseries} % Head font
  {.}      % Punctuation after head
  {0.25em}  % Space after head
  {}       % Theorem head spec

\theoremstyle{tight}

\crefname{theorem}{Thm.}{Thms.}
\crefname{lemma}{Lem.}{Lems.}
\crefname{proposition}{Prop.}{Props.}
\crefname{corollary}{Cor.}{Cors.}
\crefname{definition}{Def.}{Defs.}
\makeatletter

\def\@IEEEreftext#1#2{#1}%

\makeatother
\usepackage{aliascnt}
\usepackage{caption}
\usepackage[subrefformat=simple, labelformat=simple]{subcaption}

\usepackage{flushend}

\newcommand{\bt}[1]{\mbox{$\bf #1$}}
\def\l{\left(}
\def\r{\right)}
\newcommand{\img}{\bt x}
\newcommand{\cimg}{\hat{\bt x}}

\DeclareMathOperator*{\argmin}{arg\,min}

\newtheorem{theorem}{Theorem}[section] % Theorem numbering is tied to sections
\newtheorem{proposition}[theorem]{Proposition} % Corollary numbering follows theorems
\newaliascnt{proposition}{theorem}
\aliascntresetthe{proposition}

\newcounter{remark}

\usepackage{enumitem}

\usepackage{amsmath}

\newcommand{\gridg}{\bt L_g(\pmb{\phi})}
\usepackage{amsmath}
\usepackage{siunitx}
\newcommand{\baseL}{\bt L}

\newcommand{\bfnew}{\tilde{\pmb{\lambda}}}
\newcommand{\bfbase}{\pmb{\lambda}}

\newcommand{\diag}[1]{\mathrm{diag}(#1)}

\newcommand{\Lv}{\tilde{\bt L}(\alpha, \beta, i)}
\newcommand{\defeq}{\doteq}

\def\BibTeX{{\rm B\kern-.05em{\sc i\kern-.025em b}\kern-.08em
    T\kern-.1667em\lower.7ex\hbox{E}\kern-.125emX}}
\begin{document}

\title{INT-DTT+: Low-Complexity Data-Dependent \\ Transforms for Video Coding  
}
\author{\IEEEauthorblockN{Samuel Fernández-Menduiña$^{\color{magenta}\textsc{ sc}}$, Eduardo Pavez$^{\color{magenta}\textsc{ sc}}$, Antonio Ortega$^{\color{magenta}\textsc{ sc}}$,  \\ Tsung-Wei Huang$^{\color{magenta}\textsc{ dl}}$, Thuong Nguyen Canh$^{\color{magenta}\textsc{ dl}}$, Guan-Ming Su$^{\color{magenta}\textsc{ dl}}$, and Peng Yin$^{\color{magenta}\textsc{ dl}}$}\

\smallskip

\IEEEauthorblockA{\textit{$^{\color{magenta}\textsc{ sc}}$Department of Electrical and Computer Engineering, University of Southern California, Los Angeles, CA, US} \\
\textit{{}$^{\color{magenta}\textsc{ dl}}$Dolby Laboratories, Inc, Sunnyvale, CA, USA}}}
\maketitle

\begin{abstract}
Discrete trigonometric transforms (DTTs), such as the DCT-2 and the DST-7, are widely used in video codecs for their balance between coding performance and computational efficiency. In contrast, data-dependent transforms, such as the Karhunen–Loève transform (KLT) and graph-based separable transforms (GBSTs), offer better energy compaction but lack symmetries that can be exploited to reduce computational complexity. This paper bridges this gap by introducing a general framework to design low-complexity data-dependent transforms. Our approach builds on DTT+, a family of GBSTs derived from rank-one updates of the DTT graphs, which can adapt to signal statistics while retaining a structure amenable to fast computation. We first propose a graph learning algorithm for DTT+ that estimates the rank-one updates for rows and column graphs jointly, capturing the statistical properties of the overall block. Then, we exploit the progressive structure of DTT+ to decompose the kernel into a base DTT and a structured Cauchy matrix. By leveraging low-complexity integer DTTs and sparsifying the Cauchy matrix, we construct an integer approximation to DTT+, termed INT-DTT+. This approximation significantly reduces both computational and memory complexities with respect to the separable KLT with minimal performance loss. We validate our approach in the context of mode-dependent transforms for the VVC standard, following a rate-distortion optimized transform (RDOT) design approach. Integrated into the explicit multiple transform selection (MTS) framework of VVC in a rate-distortion optimization setup, INT-DTT+ achieves more than 3\% BD-rate savings over the VVC MTS baseline, with complexity comparable to the integer DCT-2 once the base DTT coefficients are available.
\end{abstract}


\section{Introduction}
\label{sec:intro}
Data-dependent transforms, particularly the non-separable Karhunen–Loève Transform (KLT) \cite{jain1976fast}, are theoretically optimal for linear decorrelation under some statistical assumptions \cite{effros2004suboptimality}. However, the KLT is not widely used in existing block-based video compression systems due to its lack of fast implementations \cite{bross2021overview}. Instead, most codecs rely on discrete trigonometric transforms (DTTs) \cite{puschel2008algebraic}, such as the discrete cosine transform (DCT-2) \cite{strang1999discrete} and the asymmetric discrete sine transform (ADST or DST-7) \cite{han2011jointly}. 
%Despite being suboptimal, the 
DTTs have persisted across codec generations because they \emph{approximate} the KLT for image blocks and certain prediction residuals while having arithmetic \emph{symmetries} \cite{puschel2008algebraic} that can be exploited to reduce the number of computations and memory footprint \cite{budagavi2013core, zhao2021transform}. In particular, these structural properties enable fast algorithms on floating point arithmetic based on the fast Fourier transform (FFT) \cite{puschel2008algebraic, cooley1965algorithm}. Although floating-point operations are avoided in practice to prevent encoder-decoder mismatches due to rounding or drift \cite{budagavi2013core}, integer kernels derived from DTTs—typically obtained by quantizing the original floating-point transforms \cite{budagavi2013core}—inherit their symmetries and benefit from a reduced number of arithmetic operations.

Nonetheless, using data-dependent–or learned–transforms can significantly outperform the DTTs in terms of coding efficiency \cite{zhao2021transform}. For instance, we can design data-driven mode-dependent transforms (MDTs) that leverage the statistical properties of the residuals associated with each intra-prediction mode and improve energy compaction without additional signaling cost  \cite{fan2019signal}. Similarly, different transforms can be applied for different block sizes  \cite{egilmez2020parametric} or chosen given specific perceptual image properties  \cite{fernandez-menduina_image_2023}. Given the complexity limitations of the KLT,  most learned transforms are designed to be separable  \cite{zhao2021transform}. Beyond separable KLT (sep-KLT) variants \cite{yeo2011low}, graph-based separable transforms (GBST) provide an alternative that approximates the sep-KLT while constraining the number of learnable parameters, which makes them more robust when fewer training samples are available (e.g., less common prediction modes for MDTs) \cite{egilmez2020graph}.

\begin{figure}
    \centering
    \includegraphics[width=\linewidth]{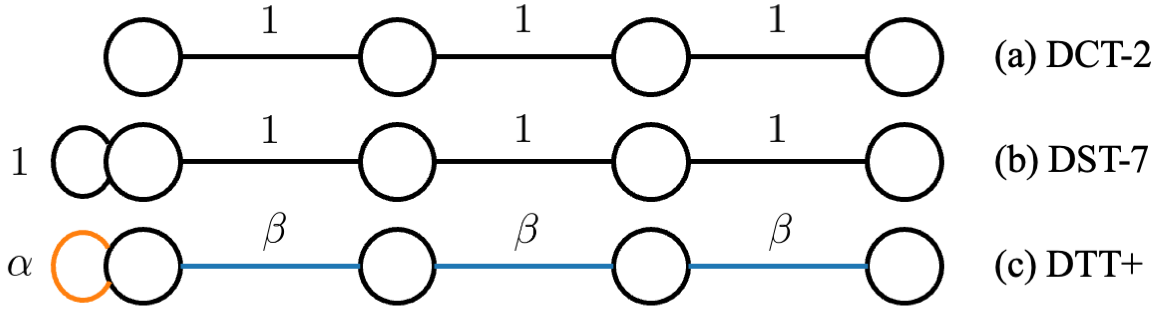}
    \caption{(a) Path graph (DCT-2), (b) path graph with unit self-loop (DST-7), and (c) DTT+ graph with parameters 
    $(\alpha, \beta) $.}
    \label{fig:path_graph}
\end{figure}

Despite complexity reductions, these separable transforms are seldom used in practice \cite{zhao2021transform}  owing to the lack of fast transforms, since the sep-KLT and the GBSTs, unlike the DTTs, do not have symmetries that can be used to reduce arithmetic operations \cite{egilmez2020parametric}. Recognizing that the lack of fast algorithms renders learned transforms impractical, this paper proposes a framework to design low-complexity data-dependent transforms via DTT+, a family of GBSTs  \cite{egilmez2020graph} derived from rank-one updates of the DTT graphs\footnote{DTTs are transforms derived from signal models with specific boundary conditions \cite{puschel2008algebraic}, which can be related to certain graphs \cite{strang1999discrete}.} \cite{fernandez-menduina2025fast}, such as the path graph (DCT-2) \cite{strang1999discrete} and the path graph with a unit self-loop (DST-7) \cite{han2011jointly}. 
This family of graphs is parametrized by two variables (cf.~\cref{fig:path_graph}) and leads to transforms with a progressive decomposition: we first apply the transform of the base graph and then multiply the result by a structured Cauchy matrix. We introduced DTT+ as a theoretical construction in \cite{fernandez-menduina2025fast}, demonstrating that these transforms have a fast algorithm in floating point arithmetic with the same asymptotic complexity as the FFT \cite{cooley1965algorithm}.  
Yet, this fast algorithm outperforms the naive matrix vector product only at large block sizes, limiting the applicability of the method in standard coding setups. Moreover, \cite{fernandez-menduina2025fast} did not address the problem of learning the best rank-one update parameters from data.

In this paper, we construct new low-complexity data-dependent transforms in two steps. First, we modify the graph learning methodology in \cite{egilmez2017graph} for the particular case of DTT+, estimating the optimal rank-one update for both row and column graphs jointly,  capturing the covariance structure of the entire block. Second, instead of quantizing the kernel directly, we implement DTT+ in integer arithmetic by exploiting progressivity and the existence of efficient integer DTTs, together with 
the structure of Cauchy matrices \cite{fernandez-menduina2025fast}, which are prone to sparsification. We call the result INT-DTT+. 
The progressive structure of INT-DTT+ is particularly useful in practical encoder implementations, where the codec searches exhaustively among all possible transform options via rate-distortion optimization (RDO) \cite{ortega_rate-distortion_1998}. Since, during RDO, the base DTT coefficients, i.e., DCT-2 and DST-7, are already computed for transform selection, the additional cost of INT-DTT+ is merely the overhead of the additional structured Cauchy matrix operations (comparable to the integer DCT-2 complexity). 
%As a result, INT-DTT+ requires a number of products comparable to the integer DCT-2 \cite{budagavi2013core} once the coefficients of the base DTTs are available. 
This structure also significantly reduces the memory required to store the transform kernel with respect to the sep-KLT. 

We test both the learned DTT+ and the corresponding INT-DTT+ in an MDT setup, extending the explicit multiple transform selection (MTS) framework of VVC for the intra-prediction residuals with an MDT. We design these transforms to complement the existing MTS kernels in VVC via a rate-distortion optimized transform (RDOT) approach  \cite{zou2013rate}. To further reduce memory complexity, we also propose a method to group similar learned transforms based on the parameters of DTT+. Results with all-intra residuals from images in the Kodak \cite{kodak1993kodak} and CLIC \cite{CLIC2022} datasets show that INT-DTT+ provides up to $3 \%$ bit-rate savings over the existing MTS setup, matching or even improving the performance of the same setup incorporating sep-KLTs, while adding a computational overhead comparable to–or even smaller than–the overhead of computing the integer DCT-2 once the DCT-2 or DST-7 coefficients are available. 

\section{Preliminaries}
\noindent\textbf{Notation.} Uppercase bold letters, such as $\bt A$, denote
matrices. Lowercase bold letters ($\bt a$) denote vectors.
The $n$th entry of $\bt a$ is $a_n$, and the $(i, j)$th entry of $\bt A$ is $A_{ij}$. Regular letters denote
scalar values. Let $\mathcal{G} = (\mathcal{V}, \mathcal{E}, \bt W, \bt V)$ be a weighted undirected graph, with vertex set $\mathcal{V}$, edge set $\mathcal{E}$, weighted adjacency matrix $\bt W$, and self-loop diagonal matrix $\bt V$. $\bt L \defeq \bt D - \bt W + \bt V$ is the generalized graph Laplacian, where $\bt D \doteq \diag{\bt 1^\top \bt W}$ is the degree matrix. Given $\bt L = \bt U \diag\bfbase \bt U^\top$, $\bt U$ is the graph Fourier transform (GFT) \cite{ortega2018graph}.

\smallskip

\noindent\textbf{DTT+.} In our previous work, we analyzed the properties of rank-one updates of the Laplacian of an arbitrary graph $\baseL\in\mathbb{R}^{n\times n}$ \cite{fernandez-menduina2025fast}. In this work, we focus on rank-one updates that add a self-loop, which we have observed empirically provide the best coding gains:
\begin{equation}
\label{eq:rank_one_base}
    \Lv \doteq \beta \, \baseL + \alpha \, \bt e_i\bt e_i^\top, \quad i = 1, \hdots n, \ \alpha \geq 0, \ \beta\geq 0,
\end{equation}
where $\bt e_i$ is the $i$th vector of the canonical basis. We let $\bt L$ be the Laplacian of a DTT graph, such as the path graph (DCT-2) \cite{strang1999discrete} and the path graph with unit self-loop (DST-7) \cite{han2011jointly} (cf.~\cref{fig:path_graph}), because their low-complexity integer versions are widely used in existing codecs. We call the GFTs associated with these Laplacians \textit{base DTTs} (i.e., DCT-2 and DST-7). In \cite{fernandez-menduina2025fast}, we proved the next result.
\begin{proposition}[Progressive decomposition \cite{fernandez-menduina2025fast}]
\label{prop:progressive}
Given $\baseL= {\bt U}\diag{\bfbase}\bt U^\top$ and $\Lv = \tilde{\bt U}\diag{\bfnew}\tilde{\bt U}^\top$, we can write:
\begin{equation}
\label{eq:fwd_transform}
    \tilde{\bt U}^\top = \diag{\bt a}\bt C(\bfnew, \beta\bfbase) \diag{\mathbf{z}} \bt U^\top.
\end{equation}
where $\bt C(\bfnew, \beta\bfbase)$ is a Cauchy matrix \cite{fernandez-menduina2025fast} such that $C_{ij} = 1/(\tilde{\lambda}_i - \beta \lambda_j)$ for $i, j = 1, \hdots, n$, $\bt z = \bt U^\top \bt e_i$, and $\bt a$ normalizes the basis.
\end{proposition}
Thus, to compute the DTT+ coefficients, we can apply the base DTT and then a Cauchy-like orthogonal matrix \cite{fernandez-menduina2025fast} to transfer between bases. We will also need the following property.
\begin{proposition}[Eigenvalue interleaving \cite{bunch1978rank}]
\label{prop:interleaving}
Provided $\alpha$ and $\beta$ are positive, the eigenvalues of the Laplacian of a DTT+ satisfy:
\begin{equation}
\label{eq:interleaving}
   \beta \lambda_1 \leq \tilde{\lambda}_1 \leq \beta\lambda_2 \leq \hdots \leq \beta\lambda_n \leq \tilde{\lambda}_n. 
\end{equation}
\end{proposition}

\smallskip

\noindent\textbf{Rate-distortion optimized transform design.}
Given $n_e$ examples $\lbrace \img_i\rbrace_{i = 1}^{n_e}$  and $n_t$ transforms $\lbrace \bt T_j\rbrace_{j = 1}^{n_t}$, we denote by $\cimg_{ij}$ the reconstruction of the $i$th block after quantization of the transform coefficients $\bt T_j^\top\img_i$. Define the index set of examples for which the $j$th transform is RD-optimal: 
\begin{equation}
\label{eq:set_rdot}
\mathcal{I}_j \doteq \left\lbrace  i \, \colon \, j = \argmin_{k = 1, \hdots, n_t} \ \norm{\img_i -   \cimg_{ik}}_2^2 + \lambda \, r(\cimg_{ik})  \right\rbrace, 
\end{equation}
for $j = 1, \hdots, n_t$, where $r(\cdot)$ denotes rate and $\lambda\geq 0$ is the Lagrangian controlling the RD trade-off \cite{ortega_rate-distortion_1998}. RDO transform (RDOT) design aims to cluster   the data while learning the transforms \cite{zou2013rate}:
\begin{equation}
    \lbrace \bt T^*_j\rbrace_{j=1}^{n_t} = \min_{\lbrace \mathbf{T}_j\rbrace^{n_t}_{j=1}} \, \sum_{j = 1}^{n_t}\, \sum_{i\in\mathcal{I}_j}\, \norm{\img_i -   \cimg_{ij}}_2^2 + \lambda \, r_i(\cimg_{ij}).
\end{equation}
We follow a Lloyd-type iterative process: we first learn the transforms given the clustered examples, and then cluster the examples assuming the transforms are fixed. The result is a set of kernels that complement each other, which is useful when using a mixed set of transforms, where some kernels are fixed to non-learnable codec transforms (e.g., DTTs). In this case, the remaining kernels are learned to extend the codec set by targeting cases where fixed transforms do not achieve good performance. 
We illustrate the RDOT setup for DTT+ in~\cref{alg:iterative_learning}.

\smallskip

\noindent\textbf{Problem formulation.}
Existing data-dependent transforms lack fast implementations, making them impractical despite superior coding performance. We address this problem by proposing a framework to design data-dependent transforms with low-complexity integer implementations, building upon our prior DTT+ work. First, we leverage a graph learning algorithm to design DTT+ from data \cite{egilmez2017graph} (\cref{sec:gl}). We rely on RDOT to obtain a set of transforms that complement modern codecs' kernels. This learned DTT+ could be implemented in integer arithmetic by quantizing each entry of the kernel, but the computational complexity would remain high. To obtain a low-complexity implementation, we 1) exploit the progressivity property of DTT+ (\cref{prop:progressive}) and the existence of low-complexity integer DTT kernels \cite{zhao2021transform}, and 2) propose a sparsification method, which can be theoretically justified from the interleaving property (\cref{prop:interleaving}), to obtain integer approximations for Cauchy matrices. We call the result INT-DTT+ (\cref{sec:integer}). Next, we detail each of these contributions.

\section{Learning DTT+ from data}
\label{sec:gl}
\begin{figure}
    \centering
    \includegraphics[width=\linewidth]{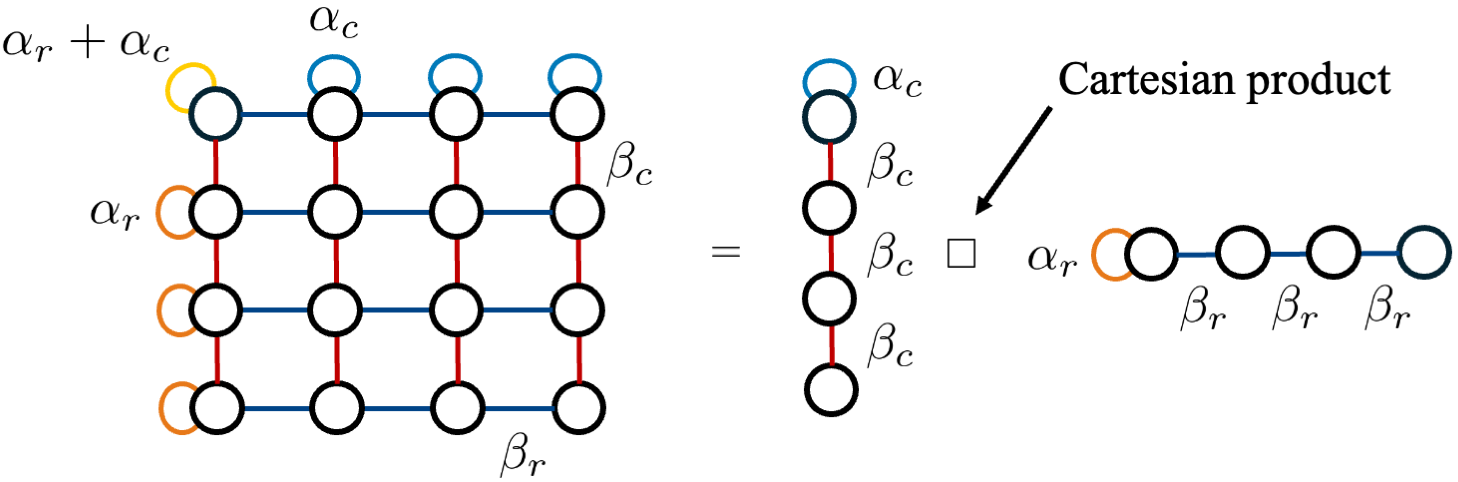}
    \caption{Graph-based model for prediction residuals as the Cartesian product of two DTT+ graphs.}
    \label{fig:cartesian}
\end{figure}

Given $n_e$ centered examples (i.e., after removing the mean of all blocks from each block) $\lbrace \bt x_i\rbrace_{i = 1}^{n_e}$, e.g., intra or inter prediction residual blocks, with sample covariance matrix $\bt S = 1/n_e \sum_{i = 1}^{n_e}\bt x_i \bt x_i^\top$, we aim to learn a DTT+ graph. To reduce computations, we focus on graphs whose eigenvector matrix can be separated into a row DTT+ and a column DTT+. To this end, we could learn the parameters $(\alpha, \beta, i)$ in \eqref{eq:rank_one_base} using rows and columns independently, but this approach would ignore interactions between rows and columns. Instead, we propose to learn both rank-one updates jointly. To achieve this goal, we first model the Laplacian of the complete block as the Cartesian product of the Laplacian of two DTT+ graphs:
\begin{equation}
    \gridg \doteq \tilde{\bt L}(\alpha_r, \beta_r, i_r)\otimes \bt I + \bt I\otimes \tilde{\bt L}(\alpha_c, \beta_c, i_c),
\end{equation}
where $\pmb{\phi} = [\alpha_r, \alpha_c, \beta_r, \beta_c, i_r, i_c]$ are parameters to learn. The transform obtained from $\gridg = \bt U_g \diag{\tilde{\pmb{\lambda}}_g}\bt U_g^\top$ is separable \cite{imrich2000product}: $\bt U_g = \bt U_r \otimes \bt U_c$, where $\tilde{\bt L}(\alpha_r, \beta_r, i_r) = \bt U_r \diag{\tilde{\pmb \lambda}_r}\bt U_r^\top$ and $\tilde{\bt L}(\alpha_c, \beta_c, i_c) = \bt U_c \diag{\tilde{\pmb \lambda}_c}\bt U_c^\top$ (\cref{fig:cartesian}). 
For $\bt L_g(\pmb{\phi})$ to remain a valid Laplacian, the parameters we learn,  $(\alpha_r, \alpha_c, \beta_r, \beta_c)$,  have to be non-negative. 
Instead of constraining the optimization problem, we use a re-parametrization trick, which yields a simpler and more interpretable solution. We remark that, since the learning algorithm runs off-line, its complexity is not critical. Now, we write
\begin{equation}
\label{eq:rank_one_base_mod}
    \Lv = \beta^2 \, \baseL + \alpha^2 \, \bt e_i\bt e_i^\top, \quad i=1, \hdots, n, \ (\alpha, \beta)\in\mathbb{R}^2.
\end{equation}
Based on this model, we seek:
\begin{equation}
\label{eq:gl_cost}
    \pmb \phi^\star = \argmin_{\pmb{\phi}} \ -\log \det(\gridg) + \tr\l \gridg\bt S\r.
\end{equation}
Unlike existing formulations that search the entire space of valid Laplacians \cite{egilmez2017graph} given a set of constraints, we focus on the set of rank-one updates to a fixed base graph. This is a mixed optimization problem because $i_r$ and $i_c$ are integers. Since only $n^2$ options are possible for these variables, we solve a continuous optimization problem for each combination of $(i_r, i_c)$ and select the optimizer of \eqref{eq:gl_cost}. Taking derivatives of \eqref{eq:gl_cost}, and exploiting rank-one structures, 
\begin{align}
\label{eq:system_dtt}
  \alpha_r \tr ((\mathbf{L}_g^{-1}(\boldsymbol{\phi}) - \bt S)(\diag{\bt e_{i_r}}\otimes \bt I))
  &= 0, \\[-0ex]
  \beta_r \tr( (\mathbf{L}_g^{-1}(\pmb\phi)-\bt S) (\mathbf{L} \otimes \mathbf{I})) 
  &= 0, \\[-0ex]  
  \alpha_c \tr( (\mathbf{L}_g^{-1}(\boldsymbol{\phi}) - \bt S)(\bt I \otimes \diag{\bt e_{i_c}}))
  &= 0, \\[-0ex]
  \beta_c \tr((\mathbf{L}_g^{-1}(\pmb\phi) - \bt S) (\mathbf{I} \otimes \mathbf{L})) 
  &= 0,
\end{align}
for $i_r, i_c = 1, \hdots, n$. We emphasize that $\bt L_g(\pmb{\phi})$ is a function of the $6$ variables in $\pmb{\phi}$. For each pair $(i_r, i_c)$, we can find a unique solution for the system of equations, $\pmb{\varphi}^\ast(i_r,i_c) = [\alpha_r, \beta_r, \alpha_c, \beta_c]$ and then compute the cost \eqref{eq:gl_cost} for $\pmb{\phi} = (i_r,i_c, \pmb{\varphi}^\ast(i_r,i_c))$. The solution to  \eqref{eq:gl_cost} is given by finding $(i_r, i_c)$ such that $(i_r,i_c, \pmb{ \varphi}^\ast(i_r,i_c))$ minimizes \eqref{eq:gl_cost}. Intuitively, we approximate the sample covariance with the covariance predicted by our model, i.e., the inverse of the DTT+ Laplacian. We find each $\pmb{\varphi}^\ast(i_r, i_c)$  using Newton's method. 

\begin{figure*}
    \centering
    \includegraphics[width=\linewidth]{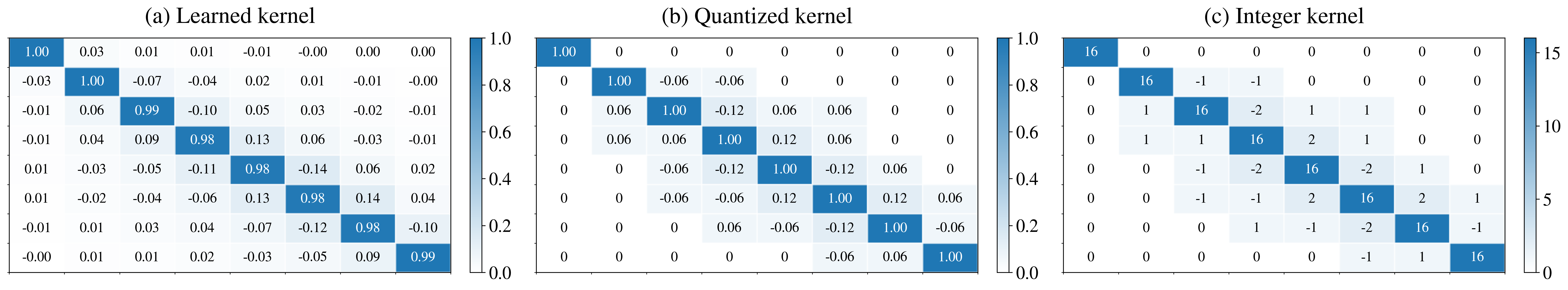}
    \caption{(a) Transition kernel between DST-7 and the DTT+ learned for the planar mode in VVC, (b) results after quantization with step $16$, and (c) its integer version, factoring out divisions by $16$. Quantization yields an integer and sparse approximation to the original kernel.}
    \vspace{-1em}
    \label{fig:numerical}
\end{figure*}

\begin{algorithm}[t]
\renewcommand{\Require}[1]{\State \textbf{Input:} #1}
\renewcommand{\Ensure}[1]{\State \textbf{Output:} #1}
\caption{DTT+ graph learning with RDOT design}
\label{alg:iterative_learning}
\begin{algorithmic}[1]
    \Require Residual blocks $\{\mathbf{x}_i\}_{i=1}^{n_e}$, number of transforms $n_t$
    \State Initialize clusters $\{\mathcal{I}_j\}_{j=1}^{n_t}$ via random partition of $\{1, \dots, n_e\}$
    \While{not converged}
    \State Find $\pmb{\phi}^*_j$ and $\bt T_j = \mathrm{eig}(\bt L_g(\pmb{\phi}^*_j))$ by solving \eqref{eq:gl_cost} using $\mathcal{I}_j$
    \State Update $\mathcal{I}_j$ as in \eqref{eq:set_rdot}, for $j=1, \hdots, n_t$
    \EndWhile
    \Ensure Learned transforms $\lbrace \mathbf{T}_j\rbrace_{j = 1}^{n_t}$
\end{algorithmic}
\end{algorithm}

\section{Integer DTT+}
\label{sec:integer}
Following the conventional approach to integer transform design \cite{budagavi2013core}, we could quantize the DTT+ kernel directly, which would have the same computational complexity as the sep-KLT. However, thanks to the progressivity property of DTT+, together with the unique symmetries of Cauchy matrices, we can obtain integer kernels that approximate the DTT+ with a reduced number of arithmetic operations. First, \cref{prop:progressive} ensures that we can compute the transform coefficients of DTT+ from the transform coefficients of the base DTT by applying a transition kernel $\bt K$, which corresponds to the Cauchy matrix. This transition kernel acts as a corrective rotation, refining the basis of the base DTT to better match the signal statistics captured by the DTT+ parameters. Since the DTTs already have low-complexity integer implementations, we can focus on $\bt K$:
\begin{equation}
\label{eq:transition}
K_{ij} = a_iz_j\big / (\tilde{\lambda}_j - \beta\lambda_i), \quad \text{for} \ \  i, j = 1, \hdots, n.
\end{equation}
Given \cref{prop:interleaving}, $\vert\tilde{\lambda}_j - \beta\lambda_i\vert$ increases as the gap $\vert i-j\vert$ increases. Since $a_i$ normalizes the rows of the transform, it tends to be roughly constant. Thus, as $\vert i-j\vert$ increases $\vert K_{ij}\vert$ decreases, so that the entries of 
$\bt K$ decay the further away we are from the diagonal (cf.~\cref{fig:numerical} for a numerical example). Now, let us write:
\begin{equation}
    \bt K = \bt K_d + \bt K_o = (\bt I + \bt K_o {\bt K_d}^{-1})\bt K_d = (\bt I + \bt F)\bt K_d,
\end{equation}
where $\bt K_d$ and $\bt K_o$ contain the diagonal and off-diagonal terms, respectively. We write $\bt K$ in terms of $\bt F \doteq \bt K_o{\bt K_d}^{-1}$ because $\bt F$ can be sparsified more aggressively than $\bt K_o$. Now, we apply bit-level scalar quantization with clipping by computing:
\begin{equation}
\label{eq:quantization}
\bt K_{dq} \doteq \mathrm{round}(p_d \, \bt K_d) / p_d \ \ \text{and} \ \ \bt F_q \doteq \mathrm{round}(p_{\mathsf f}\, \bt F) / p_{\mathsf f},
\end{equation}
and clipping the results to obtain $\bt F_q'$ and $\bt K_{dq}'$, with coefficients in a given bit-depth precision. We choose $p_{\mathsf{f}}$ and $p_d$, parameters that control the precision of the integer approximation, so that the division can be implemented using binary shifts. Moreover, we let $p_d$ be larger than $p_{\mathsf f}$ to punish the out-of-diagonal terms more aggressively, which reinforces the already existing decay. We measure the approximation quality by means of orthogonality, closeness to the original kernel, and norm, as in HEVC/VVC \cite{budagavi2013core}. After quantization, we fine-tune the entries of the kernel in the range $\pm 1$ to optimize these three metrics, following the strategy used for the DTTs in HEVC/VVC \cite{budagavi2013core}.

The forward transform is summarized in \cref{alg:transform_algorithm}. The inverse transform is the transpose matrix, and has similar complexity. During the RDO, we assume the coefficients for the base DTT (DCT-2 or DST-7) are already computed using integer kernels. Once the DTT coefficients are available, we 1) multiply by the quantized diagonal term $\bt K_{dq}'$, 2) multiply the result by the sparsified kernel $\bt F_q'$, and 3) add up the results of the two previous steps. The next section analyzes the performance in a realistic coding scenario.

\begin{algorithm}[t]
\renewcommand{\Require}[1]{\State \textbf{Input:} #1}
    \caption{INT-DTT+, forward transform}
    \label{alg:transform_algorithm}
    \begin{algorithmic}[1]
        \Require Image block $\img_i$, INT-DTT+ matrices $\bt K'_{dq}$ and $\bt F'_q$.        
        \State Compute base DTT coefficients, $\bt y_i = \bt U^\top \bt x_i$.
        \State Multiply by diagonal matrix $\bt z_i = \bt K'_{dq}\bt y_i$. 
        \State Return $\bt q_i = \bt z_i + \bt F'_q\bt z_i$ 
    \end{algorithmic}
\end{algorithm}

\section{Experimental evaluation}
We consider mode-dependent transforms for planar, DC, and angular VVC intra-prediction residuals. For training, we consider images from the CLIC testing set \cite{CLIC2022} (from $878\times 2048$ to $2048 \times 2048$ pixels) and Kodak \cite{kodak1993kodak} ($512\times 768$ pixels). For testing, we consider images from the CLIC validation set (from $878\times 2048$ to $2048 \times 2048$ pixels). We obtain residuals by running the codec in all-intra configuration and  selecting only the residual blocks (before quantization) where $8\times 8$, $16\times 16$, or $32\times 32$ is the optimal RD choice.

We encode the prediction residuals using VVC's explicit multiple transform selection (MTS) setup. DTT+, INT-DTT+, and the sep-KLT are added to the MTS candidate set and compete during RDO. The transform index is signaled using VVC's existing MTS syntax, while the transform parameters themselves are fixed offline during codec design and indexed by prediction mode. DTT+ and sep-KLT are quantized to $8$-bits precision. We consider RDOT design with the VVC-MTS kernels in all cases. We speed up RDOT by using the $\ell_1$ norm of the transform coefficients as a bit-rate proxy \cite{pakiyarajah2025joint}. We repeat RDOT steps until the decrease in RD cost is $1\%$. For training, we use the same number of samples for all modes. We use a deadzone quantizer \cite{zhao2021transform} and  CABAC \cite{wiegand_overview_2003}, with PSNR as the distortion metric.

\begin{figure}
    \centering
    \includegraphics[width=\linewidth]{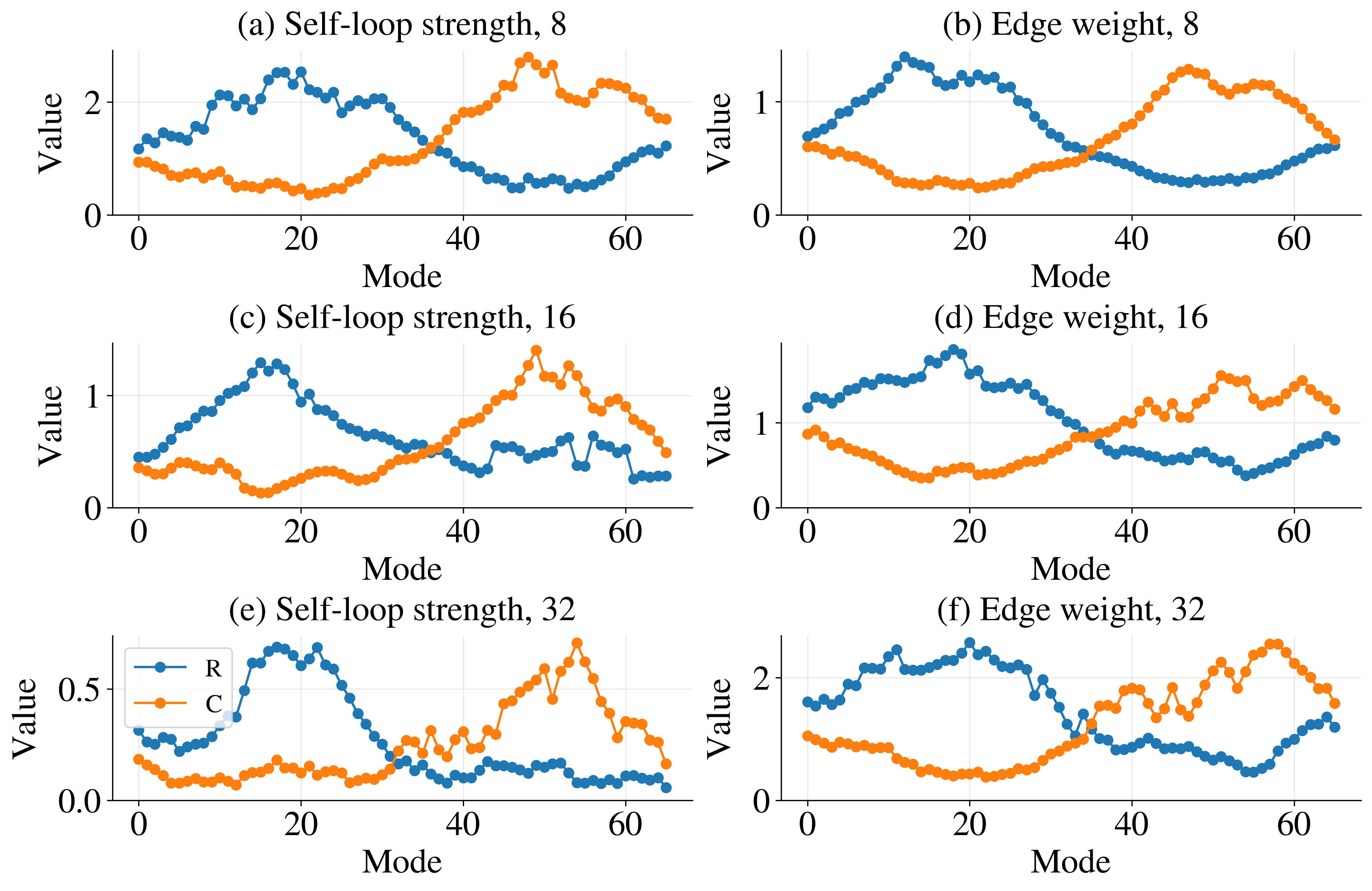}
    \caption{Learned weights for rows (R) and columns (C), with $8\times 8$, $16\times 16$, and $32\times 32$ blocks. We observe spatial consistency.}
    \label{fig:line_plots}
\end{figure}

\textbf{Graph learning.} The base graph for learning is the path graph (\cref{fig:path_graph}(a)). We find the DTT+ parameters for each mode by solving \eqref{eq:gl_cost} in a RDOT setup. We show the learned parameters for each intra-prediction mode for  $8\times 8$, $16\times 16$, and $32\times 32$ blocks in \eqref{eq:gl_cost}, estimated from $2000$ blocks for each mode (Fig.~\ref{fig:line_plots}). The optimal position for the self-loop is always on the first node, i.e., on average, the prediction is best at the boundary pixels. The parameters $(\alpha_r, \alpha_c, \beta_r, \beta_c)$ are similar for modes corresponding to contiguous prediction angles, which shows the statistical similarity of their residuals. $\alpha_r$ peaks for horizontal prediction (mode 18), while $\alpha_c$ peaks for vertical prediction (mode 50). Increasing block size decreases self-loop weights and increases edge weights \cite{egilmez2020parametric}.

\begin{figure}[t]
    \centering
    \includegraphics[width=\linewidth]{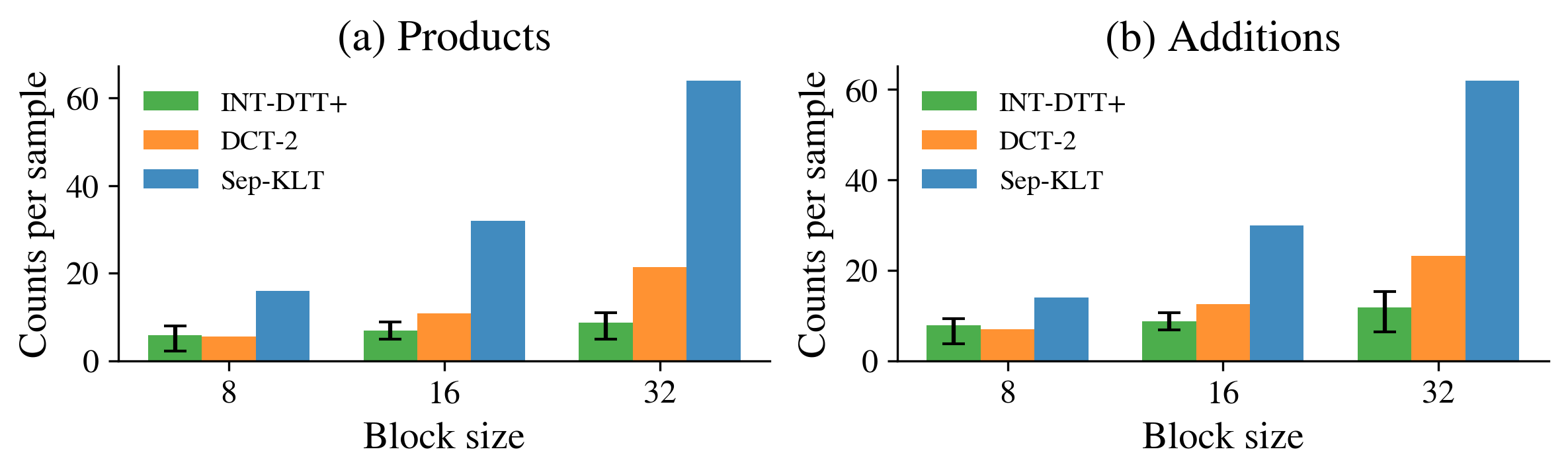}
    \caption{Operation count for the forward integer transforms. For INT-DTT+, we assume the coefficients of the base DTT are available, i.e., RDO scenarios, where the DTTs are almost always computed, and show median, maximum, and minimum across all modes. INT-DTT+ compares to the integer DCT-2 in complexity.}
    \vspace{-1em}
    \label{fig:num_prods}
\end{figure}

\textbf{INT-DTT+.} From Weyl's inequality \cite{golub2013matrix}, the eigenvalue gap between base DTT and DTT+ in \cref{prop:interleaving}, which controls coefficient decay in the Cauchy matrix, depends on the gap between the self-loop weights on each graph. Thus, we choose the DCT-2 (self-ltoop $0$) as the base transform when the learned self-loop weight is smaller than $0.5$ and the DST-7 (self-loop $1$) otherwise. We count arithmetic operations by grouping terms row-wise and ignoring products by $1$. 
Following \eqref{eq:quantization}, we use $8$ bits with $p_d = 128$ for the diagonal matrix $\bt D$, which is typical for integer transforms \cite{budagavi2013core}, and $3$ bits with $p_{\sf f} = 4$ for the out-of-diagonal matrix $\bt F$, which yields experimentally a good trade-off between complexity and coding performance. \cref{fig:num_prods} shows the operation count to obtain the final transform coefficients. We assume RDO encoding, i.e., the encoder first computes the coefficients for the base DTTs, and then INT-DTT+ applies the sparse transition kernel over the DTT coefficients. Since the base DTT operations have to be performed anyway,  we exclude them from the INT-DTT+ complexity. For $8\times 8$ blocks, INT-DTT+ needs a similar number of products as the integer DCT-2, with further reductions for larger blocks. The total count is less than twice the DCT-2 count when computing INT-DTT+ directly to or from pixel-domain residuals, e.g., for decoder-side computations.

\textbf{Coding experiments.} We test $5000$ $8\times 8$ intra-prediction residuals per angular mode. The baseline is VVC explicit MTS. We report the averaged BD-rate \cite{bjontegaard_calculation_2001} across all modes for multiple training set sizes (\cref{tab:metrics_results}). Since our model includes only two parameters, DTT+ and INT-DTT+ are more robust to smaller training set sizes than the sep-KLT. We also consider different block sizes with $2000$ training samples per mode (\cref{tab:block_size}). While the sep-KLT requires learning $n^2$ parameters, DTT+ involves two parameters regardless of the block size. Since the number of training samples is fixed but the number of parameters for the sep-KLT increases, overfitting may lead to worse performance in the test set. This problem is mitigated by DTT+ since the number of parameters is small and independent of the block size. Thus,  DTT+ may outperform the sep-KLT in terms of BD-rate. INT-DTT+ introduces minimal loss and reduces arithmetic computations.

\textbf{Weight clustering.} Since integer DTTs are available in the codec, INT-DTT+ needs only the integer version of the transition kernels, reducing the storage requirements over the sep-KLT and DTT+. Nonetheless, if memory is an issue, storing a transform per mode might be impractical. We can reduce the number of kernels by clustering the estimated self-loop weights (cf.~\cref{fig:line_plots}) via k-means \cite{macqueen1967some} applied jointly to the row and column weights. Weight clustering can identify similar transforms regardless of the relationships between modes, improving over grouping transforms based on the prediction angle.  We repeat our coding experiments with $2000$ training samples while varying the number of MDT kernels (\cref{tab:grouping}). \cref{tab:memory} shows the memory required by INT-DTT+, showing reductions of $66\, \%$ for $8\times 8$, $82\, \%$ for $16\times 16$, and $94\,\%$ for $32\times 32$ over the sep-KLT.

        \begin{table}[t]
        \centering
        \renewcommand{\baselinestretch}{0.5}      
        \setlength{\tabcolsep}{16pt} % Default value: 6pt
        \begin{tabular}{@{}lcccc@{}}
            \toprule \textbf{Samples} & \textbf{sep-KLT} & \textbf{DTT+} & \textbf{INT-DTT+}\\
            \midrule
             $500$ & $-2.70$ & $\mathbf{-3.06}$ & $-3.01$ \\
             $1000$ &  $-2.99$ & $\mathbf{-3.08}$ & $-3.04$ \\
             $2000$ &  $\mathbf{-3.21}$ & $-3.12$ & $-3.06$ \\
             $4000$ &  $\mathbf{-3.25}$ & $-3.13$ & $-3.09$ \\
            \bottomrule
        \end{tabular}
        \caption{Average BD-rate savings ($\%$) with respect to VVC MTS considering different training set sizes. Lower is better. DTT+ and INT-DTT+ are more robust to reducing the size of the training set. INT-DTT+ introduces minimal performance loss.}
        \label{tab:metrics_results}
    \end{table}

   \begin{table}[t]
        \centering
        \renewcommand{\baselinestretch}{0.5}
        \setlength{\tabcolsep}{16pt} % Default value: 6pt
        \begin{tabular}{@{}lcccc}
            \toprule \textbf{Size} & \textbf{sep-KLT} & \textbf{DTT+} & \textbf{INT-DTT+}\\
            \midrule
             $8\times 8$ & $\mathbf{-3.21}$ & $-3.12$ & $-3.06$ \\
             $16\times 16$ &  $-3.60$ & $\mathbf{-3.64}$ & $-3.46$ \\
             $32\times 32$ &  $-3.72$ & $\mathbf{-3.96}$ & $-3.75$ \\
            \bottomrule
        \end{tabular}
        \caption{Averaged BD-rate savings ($\%$) with respect to VVC-MTS. Lower is better. DTT+ yields further coding gains for larger blocks.}
        \label{tab:block_size}
    \end{table}

   \begin{table}[t]
        \centering
        \setlength{\tabcolsep}{6.5pt}
        \renewcommand{\baselinestretch}{0.7}        
        \begin{tabular}{@{}lccccc}
            \toprule \textbf{Method} & $3$ & $4$ & $5$ & $6$ & $7$ \\
            \midrule
             sep-KLT &  $\mathbf{-2.92}$ & $-3.01$ & $-3.06$ & $-3.08$ & $\mathbf{-3.12}$\\
             DTT+ & $-2.89$ & $-2.96$ & $\mathbf{-3.08}$ & $\mathbf{-3.13}$ & $-3.14$\\
             INT-DTT+ &  $-2.85$ & $\mathbf{-3.02}$ & $-3.04$ & $-3.06$ & $-3.08$\\
            \bottomrule
       \end{tabular}
        \caption{Averaged BD-rate savings ($\%$) with respect to VVC MTS for $8\times 8$ blocks as a function of the number of kernels after grouping. Lower is better. DTT+ and INT-DTT+ use our weight-clustering; sep-KLT uses angle-based grouping. $6$ kernels match the performance of having one transform for each of the $66$ modes.}
        \label{tab:grouping}
    \end{table}

   \begin{table}[t]
        \centering
        \renewcommand{\baselinestretch}{0.7}    
        \setlength{\tabcolsep}{6pt} % Default value: 6pt
        \begin{tabular}{@{}lccccc|c}
            \toprule \textbf{Size} & $3$ & $4$ & $5$ & $6$ & $7$ & 
            $1$ sep-KLTs\\
            \midrule
             $8\times 8$ & $1152$ & $1536$ & $1976$ & $2384$ & $2784$ & $1024$\\
             $16\times 16$ &  $1552$ & $2112$ & $2616$ & $3152$ & $3640$ & $4096$ \\
             $32\times 32$ &  $2176$ & $3408$ & $4520$ & $5528$ & $6464$ & $16384$\\
            \bottomrule
        \end{tabular}
        \caption{INT-DTT+ memory usage (bits) with number of kernels after clustering. For sep-KLT, total memory scales linearly with the number of kernels.}
        \label{tab:memory}
    \end{table}

\section{Conclusion}
\label{sec:majhead}
This paper proposes a framework to obtain low-complexity integer data-dependent kernels by leveraging transforms derived from rank-one updates of the DTT graphs (DTT+). We first propose a graph learning method to find the best rank-one update from data samples. Then, by exploiting progressivity, the existence of integer DTT kernels, and the structure of Cauchy matrices, we provide a low-complexity integer implementation of DTT+, or INT-DTT+. We tested our ideas in the explicit MDT setup of VVC. We used intra-prediction residuals to learn the weights of DTT+, combined with a rate-distortion optimized transform (RDOT) setup. Results show coding gains against VVC MTS of more than $3\%$, with computational complexity similar to the integer DCT-2 once the base DTT coefficients are available. Future work will consider inter-prediction modes, hardware-aware runtime evaluations, and other codecs.

\bibliographystyle{IEEEtran}
\bibliography{bibfile}
\end{document}